\documentclass[twocolumn]{aastex631}  

\usepackage{amsmath}
\usepackage{xcolor}
\usepackage{bm}
\usepackage{graphicx}
\usepackage{txfonts}
\usepackage{verbatimbox}
\shorttitle{Simulating the detection of the global 21~cm signal with MIST}
\shortauthors{Monsalve et al.}

\begin{document}

\title{\large Simulating the Detection of the Global 21~cm Signal with MIST for Different Models of the Soil and Beam Directivity}

\author[0000-0002-3287-2327]{Raul A. Monsalve}
\affiliation{Space Sciences Laboratory, University of California, Berkeley, CA 94720, USA}
\affiliation{School of Earth and Space Exploration, Arizona State University, Tempe, AZ 85287, USA}
\affiliation{Departamento de Ingenier\'ia El\'ectrica, Universidad Cat\'olica de la Sant\'isima Concepci\'on, Alonso de Ribera 2850, Concepci\'on, Chile}

\correspondingauthor{Raul A. Monsalve}
\email{raul.monsalve@berkeley.edu}

\author[0000-0002-7971-3390]{Christian H. Bye}
\affiliation{Department of Astronomy, University of California, Berkeley, CA 94720, USA}

\author[0000-0001-6903-5074]{Jonathan L. Sievers}
\affiliation{Department of Physics and Trottier Space Institute, McGill University, Montr\'eal, QC H3A 2T8, Canada}
\affiliation{School of Chemistry and Physics, University of KwaZulu-Natal, Durban, South Africa}

\author[0009-0008-9653-6104]{Vadym Bidula}
\affiliation{Department of Physics and Trottier Space Institute, McGill University, Montr\'eal, QC H3A 2T8, Canada}

\author[0000-0001-8468-9391]{Ricardo Bustos}
\affiliation{Departamento de Ingenier\'ia El\'ectrica, Universidad Cat\'olica de la Sant\'isima Concepci\'on, Alonso de Ribera 2850, Concepci\'on, Chile}

\author[0000-0002-4098-9533]{H. Cynthia Chiang}
\affiliation{Department of Physics and Trottier Space Institute, McGill University, Montr\'eal, QC H3A 2T8, Canada}
\affiliation{School of Mathematics, Statistics, \& Computer Science, University of KwaZulu-Natal, Durban, South Africa}

\author[0009-0002-9727-8326]{Xinze Guo}
\affiliation{Space Sciences Laboratory, University of California, Berkeley, CA 94720, USA}

\author[0009-0003-3736-2080]{Ian Hendricksen}
\affiliation{Department of Physics and Trottier Space Institute, McGill University, Montr\'eal, QC H3A 2T8, Canada}

\author[0009-0005-1658-6071]{Francis McGee}
\affiliation{Department of Physics and Trottier Space Institute, McGill University, Montr\'eal, QC H3A 2T8, Canada}

\author{F. Patricio Mena}
\affiliation{National Radio Astronomy Observatory, Charlottesville, VA 22903, USA}
\affiliation{Departamento de Ingenier\'ia El\'ectrica, Universidad de Chile, Santiago, Chile}

\author[0000-0003-4242-2685]{Garima Prabhakar}
\affiliation{Space Sciences Laboratory, University of California, Berkeley, CA 94720, USA}

\author[0000-0002-2924-9278]{Oscar Restrepo}
\affiliation{Facultad de Ingenier\'ia, Universidad ECCI, Bogot\'a, 111311, Colombia}
\affiliation{Departamento de Ingenier\'ia El\'ectrica, Universidad de Chile, Santiago, Chile}

\author[0000-0003-1602-7868]{Nithyanandan Thyagarajan}
\affiliation{Commonwealth Scientific and Industrial Research Organisation (CSIRO), Space \& Astronomy, P. O. Box 1130, Bentley, WA 6102, Australia}

\begin{abstract} 
The Mapper of the IGM Spin Temperature (MIST) is a new ground-based, single-antenna, radio experiment attempting to detect the global $21$~cm signal from the Dark Ages and Cosmic Dawn. A significant challenge in this measurement is the frequency-dependence, or chromaticity, of the antenna beam directivity. MIST observes with the antenna above the soil and without a metal ground plane, and the beam directivity is sensitive to the electrical characteristics of the soil. In this paper, we use simulated observations with MIST to study how the detection of the global $21$~cm signal from Cosmic Dawn is affected by the soil and the MIST beam directivity. We simulate observations using electromagnetic models of the directivity computed for single- and two-layer models of the soil. We test the recovery of the Cosmic Dawn signal with and without beam chromaticity correction applied to the simulated data. We find that our single-layer soil models enable a straightforward recovery of the signal even without chromaticity correction. Two-layer models increase the beam chromaticity and make the recovery more challenging. However, for the model in which the bottom soil layer has a lower electrical conductivity than the top layer, the signal can be recovered even without chromaticity correction. For the other two-layer models, chromaticity correction is necessary for the recovery of the signal and the accuracy requirements for the soil parameters vary between models. These results will be used as a guideline to select observation sites that are favorable for the detection of the Cosmic Dawn signal.
\end{abstract}



\section{Introduction}
The sky-averaged, or global, redshifted $21$~cm signal from neutral hydrogen in the intergalactic medium (IGM) is expected to reveal how the Universe evolved before and during the formation of the first stars \citep{furlanetto2006a,pritchard2008}. Models for this signal primarily consist of two absorption features in the radio spectrum: one at $\nu\lesssim50$~MHz, produced during the Dark Ages \citep{mondal2023}, and the second one in the range $50\;\mathrm{MHz}\lesssim \nu \lesssim 150$~MHz, due to the appearance of the first stars at Cosmic Dawn \citep{tozzi2000,furlanetto2006b}. 

The Mapper of the IGM Spin Temperature (MIST) is a new experiment designed to measure the global $21$~cm signal. The MIST instrument is a ground-based, single-antenna, total-power radiometer measuring in the range $25$--$105$~MHz, which encompasses the Dark Ages and Cosmic Dawn. The antenna used by MIST is a horizontal blade dipole, which operates without a metal ground plane in order to avoids systematics from ground plane resonances \citep{bradley2019} and edge effects \citep{mahesh2021,rogers2022,spinelli2022}. MIST runs on $12$~V batteries, has a power consumption of only $17$~W, and is very compact, with all the electronics and batteries housed in a single receiver box located under the antenna. The low power consumption and compactness enable MIST to be easily transported to multiple observation sites. The instrument design and initial performance are described in \citet{monsalve2023}.

One of the greatest instrumental challenges in detecting the $21$~cm signal is the ``beam chromaticity'', that is, the change in the antenna beam directivity as a function of frequency. The beam chromaticity changes the spatial weighting of the sky brightness temperature distribution across frequency. This change in spatial weighting produces structure in the spectrum that could mask or mimic the global $21$~cm signal \citep[e.g.,][]{vedantham2014,bernardi2015,mozdzen2016,sims2023}. Furthermore, the beam chromaticity of ground-based instruments is sensitive to the properties of the soil, which increases the complications and uncertainties in the modeling of the observations \citep[e.g.,][]{mahesh2021,singh2022,spinelli2022}. For MIST, which operates without a ground plane, understanding the influence of the soil on the beam chromaticity and the detectability of the $21$~cm signal is critical.

This paper studies the extraction of the Cosmic Dawn absorption feature from simulated observations with MIST. Specifically, we address the following questions: (1) How does the MIST beam chromaticity bias the $21$~cm parameter estimates? (2) How does this bias depend on the electrical properties of the soil? (3) To what extent could this bias be reduced by correcting the data for beam chromaticity? And (4) how accurately do we need to know the soil parameters when computing the chromaticity correction? To address these questions, we use the nine models of the MIST beam directivity introduced in \citet{monsalve2023}. These directivity models were produced through electromagnetic simulations that incorporated different models for the soil.

\section{Analysis}

\begin{table}
\caption{Soil models used in the FEKO simulations of MIST. These models were introduced in \citet{monsalve2023}. Five of these models (``nominal'' and \url{1L_xx}) are single-layer models and four (\url{2L_xx}) are two-layer models. The layers are characterized in terms of their electrical conductivity ($\sigma$) and relative permittivity ($\epsilon_r$). In the two-layer models, the thickness of the top layer is $L=1$~m.}             
\label{table_soil_parameters}      
\centering                          
\begin{tabular}{l c c c c c}        
\hline 
\\
Model & \# Layers & $\sigma_1$~[Sm$^{-1}$] & $\epsilon_{r1}$ & $\sigma_2$~[Sm$^{-1}$] & $\epsilon_{r2}$\\ 
\hline                        
nominal    & 1 & $0.01$ &  $6$ & & \\
\hline
\verb~1L_c+~       & 1 & $0.1$ &  $6$ & & \\
\verb~1L_c-~       & 1 & $0.001$ &  $6$ & &\\
\verb~1L_p+~       & 1 & $0.01$ &  $10$ & & \\
\verb~1L_p-~       & 1 & $0.01$ &  $2$ & & \\
\hline
\verb~2L_c+~       & 2 & $0.01$ &  $6$ & $0.1$ &  $6$ \\
\verb~2L_c-~       & 2 & $0.01$ &  $6$ & $0.001$ &  $6$ \\
\verb~2L_p+~       & 2 & $0.01$ &  $6$ & $0.01$ &  $10$ \\
\verb~2L_p-~       & 2 & $0.01$ &  $6$ & $0.01$ &  $2$ \\
\hline                                   
\end{tabular}
\end{table}

\subsection{Simulated Observations}

We simulate the sky-averaged antenna temperature spectrum that MIST would measure from the McGill Arctic Research Station (MARS) in the Canadian High Arctic ($79.37980^{\circ}$~N, $90.99885^{\circ}$~W). MARS is one of the sites from where MIST has already conducted observations \citep{monsalve2023}. We simulate the observations in the restricted frequency range $45$--$105$~MHz to focus on the absorption feature from the Cosmic Dawn. The observations have a frequency resolution of $1$~MHz.

The noiseless, time-dependent antenna temperature spectrum, $T_S$, is simulated as

\begin{equation}
T_S(\nu, t) = T_{21}(\nu) + T_{fg}(\nu, t),
\label{equation_temperature_spectrum_LST}
\end{equation}

\noindent where $T_{21}$ is the Cosmic Dawn $21$~cm absorption feature, $T_{fg}$ is the contribution from the astrophysical foreground, $\nu$ is frequency, and $t$ is time. Noise is added to the simulated observations as described in Section~\ref{section_lst_average_noise}.

We simulate the time-independent $21$~cm signal as a Gaussian, that is,

\begin{equation}
T_{21}(\nu) = a_{21} \exp\Biggl\{-\frac{1}{2}\left[\left(\frac{\sqrt{8\log2}}{w_{21}}\right)\left(\nu-\nu_{21}\right)\right]^2\Biggr\},
\label{equation_global_model}
\end{equation}

\noindent where $a_{21}$, $\nu_{21}$, and $w_{21}$ are the Gaussian amplitude, center frequency, and full width at half maximum (FWHM), respectively. We have chosen a phenomenological model for the $21$~cm signal because here we focus on the detectability of the signal instead of on its physical interpretation, similarly to previous works \citep[e.g.,][]{bernardi2016,monsalve2017,bowman2018,spinelli2019,spinelli2022,anstey2023}. Existing physical models differ in their astrophysical parameters, values, and assumptions, as well as in their computational implementation, but the shape of most of them can be well approximated by a Gaussian \citep[e.g.,][]{furlanetto2006b,pritchard2008, mesinger2011, mirocha2014, cohen2017, mirocha2021, munoz2023}. A Gaussian analytical model can be quickly evaluated in a likelihood computation and has parameters that are easy to interpret. In our simulations, the input Gaussian parameter values are $a_{21}=-200$~mK, $\nu_{21}=80$~MHz, and $w_{21}=20$~MHz. These values are consistent with standard physical models \citep{furlanetto2006a}, and our choices for the center frequency and width are close to the values reported by the EDGES experiment \citep{bowman2018}.

The time-dependent foreground contribution to Equation~\ref{equation_temperature_spectrum_LST} is computed as  

\begin{align}
T_{fg}(\nu, t) = \frac{\int_0^{2\pi} \int_{0}^{\pi/2} T_{GSM}(\theta,\phi,\nu, t)D(\theta,\phi,\nu) \sin\theta d \theta d\phi}{\int_0^{2\pi} \int_{0}^{\pi/2} D(\theta,\phi,\nu) \sin\theta d \theta d\phi}.\label{equation_convolution}
\end{align}

\noindent Here, $D$ is a model for the antenna beam directivity, which we discuss in Section~\ref{section_directivity}; $\theta$ and $\phi$ are the zenith and azimuth angles, respectively, of both the antenna and the sky; and $T_{GSM}$ is a model for the spatially dependent foreground brightness temperature distribution, for which we use the Global Sky Model \citep[GSM;][]{de_oliveira_costa2008,price2016}. We compute $T_{fg}$ across $24$~hr of local sidereal time (LST) with a cadence of $6$~minutes. In the computations, the excitation axis of the dipole antenna is aligned north-south and the horizontal blades of the antenna are perfectly level. 

The simulated observations assume perfect receiver calibration and correction of losses, in particular radiation loss, balun loss, and ground loss. Effects from the ionosphere \citep{vedantham2014}, polarized diffuse foreground emission \citep{spinelli2019}, and mountains in the horizon \citep{bassett2021,pattison2023} are assumed to be perfectly calibrated out.

\subsection{Beam Directivity}
\label{section_directivity}
We simulate sky observations with the nine beam directivity models introduced in \citet{monsalve2023}. The directivity models were obtained from electromagnetic simulations with FEKO\footnote{\url{https://www.altair.com/feko}} that incorporated different models for the soil. The characteristics of the soil models are shown in Table~\ref{table_soil_parameters}, reproduced from \citet{monsalve2023}. In the FEKO simulations, the soil model extends to infinity in the horizontal direction and in depth. Five of the soil models (``nominal'' and \verb~1L_xx~) are single-layer models. These models intend to mimic the optimistic scenario of a soil that, from the point of view of the antenna, is effectively homogeneous and well characterized in terms of only two parameters: its electrical conductivity ($\sigma_1$) and relative permittivity ($\epsilon_{r1}$). The other four soil models (\verb~2L_xx~) are two-layer models, in which the top layer has a thickness $L=1$~m and the bottom layer extends to infinite depth. Except for one parameter of the bottom layer (either $\sigma_2$ or $\epsilon_{r2}$), the parameters of the two-layer models take the same values as in the nominal model. The two-layer models are used to study the more realistic scenario in which the characteristics of the soil change below a certain depth. An example of this scenario is the soil at MARS during the summer, which consists of an unfrozen top layer and a permanently frozen, or ``permafrost,'' bottom layer \citep[e.g.,][]{pollard2009,wilhelm2011}. 

For the nominal model we use $\sigma_1=0.01$~Sm$^{-1}$ and $\epsilon_{r1}=6$. In the other models, the conductivity of the single- or bottom-layer is changed to $0.1$ and $0.001$~Sm$^{-1}$, and the relative permittivity is changed to $10$ and $2$. These values are motivated by values reported for geological materials that we could encounter at our observation sites. For instance, snow, freshwater ice, and permafrost have $\sigma$ and $\epsilon_r$ in the ranges $0$--$0.01$~Sm$^{-1}$ and $2$--$6$, respectively. For sand, silt, and clay, $\sigma$ and $\epsilon_{r}$ have a strong sensitivity to moisture and span the wide ranges $0$--$1$~Sm$^{-1}$ and $2$--$40$ \citep{reynolds2011}. Our parameter values are consistent with the values reported by \citet{sutinjo2015} for the soil at the Inyarrimanha Ilgari Bundara, the CSIRO Murchison Radio-astronomy Observatory, with different moisture levels. Our values are also consistent with the soil measurements done at the Owens Valley Radio Observatory for dry and wet conditions, presented in \citet{spinelli2022}.

As shown in \citet{monsalve2023}, for single-layer models the directivity has a smooth frequency evolution. Two-layer soil models, in contrast, produce ripples in the directivity as a function of frequency, which are expected to complicate the extraction of the $21$~cm signal. The period of the ripples depends on the depth of the interface between the two layers. The amplitude of the ripples depends on the difference in parameter values between the two layers as well as on the sign of the change. Among our two-layer models, \verb~2L_c-~ produces the smallest ripples, followed by \verb~2L_p+~, \verb~2L_p-~, and \verb~2L_c+~. 

The nine soil models considered in this paper are used to provide initial intuition about the impact of the soil properties on the detection of the Cosmic Dawn signal with MIST. We leave to future work an in-depth study of soil effects with a wide range of models. Additional models to consider would include (1) two-layer models with different top-layer thicknesses; (2) models with more than two layers (which were discussed in \citet{spinelli2022} in the context of the LEDA experiment); and (3) models with frequency dependence in the parameters.

\subsection{Chromaticity Correction}
The simulated observations computed with Equation~\ref{equation_convolution} are affected by the chromaticity of the beam. In this paper, we study the extraction of the Cosmic Dawn signal from the observations affected by this chromaticity, as well as after applying a correction to remove this effect. The chromaticity correction used is the one introduced in \citet{monsalve2017}:

\begin{align}
C(\nu, t) = &\left(\frac{\int_0^{2\pi} \int_{0}^{\pi/2} T_{GSM}(\theta,\phi,\nu, t)D_c(\theta,\phi,\nu) \sin\theta d \theta d\phi}{\int_0^{2\pi} \int_{0}^{\pi/2} D_c(\theta,\phi,\nu) \sin\theta d \theta}\right) \nonumber \\ &\times\left(\frac{\int_0^{2\pi} \int_{0}^{\pi/2} T_{GSM}(\theta,\phi,\nu, t)D_c(\theta,\phi,\nu_r) \sin\theta d \theta d\phi}{\int_0^{2\pi} \int_{0}^{\pi/2} D_c(\theta,\phi,\nu_r) \sin\theta d \theta}\right)^{-1}. \label{equation_chromaticity}
\end{align}

\noindent Here, $T_{GSM}$ is the foreground brightness temperature distribution from the GSM, $D_c$ is a model for the beam directivity used for chromaticity correction, and $\nu_r$ is the reference frequency for chromaticity correction. We use $\nu_r=75$~MHz, which is the center of our frequency range. Applying the chromaticity correction involves dividing by $C$ the data produced by Equation~\ref{equation_convolution}. In Equation~\ref{equation_chromaticity}, the numerator (denominator) corresponds to the foreground antenna temperature measured by a frequency-dependent (-independent) beam directivity. Therefore, $C$ is the factor by which the frequency-dependent directivity modifies the sky-averaged foreground spectrum that would be measured if the directivity were frequency independent and equal to $D_c(\theta,\phi,\nu_r)$. Note that $C$ is computed assuming that there is no $21$~cm signal. We account for the effect of the chromaticity correction on the $21$~cm signal at the signal extraction stage (Section~\ref{section_modeling}).

We study the extraction of the signal with three types of chromaticity correction: (1) no correction (NC), which is equivalent to making $C=1$; (2) imperfect correction, where $D_c$ in Equation~\ref{equation_chromaticity} has been computed with errors in the soil model relative to $D$ in Equation~\ref{equation_convolution}; and (3) perfect correction (PC), where $D_c=D$. In this paper we only address complications related to beam chromaticity. Therefore, $C$ is always computed with a perfect model for the foreground, that is, the GSM, also used to simulate the observations with Equation~\ref{equation_convolution}.

\subsection{Errors in Soil Parameters}
We implement the second type of chromaticity correction introduced above as follows. For each of the soil models used to produce the simulated observations, we compute several $D_c$ that are imperfect due to errors in the soil model assumed for chromaticity correction. In the assumed soil models, the number of layers is the same as in the input soil models, that is, they are both single layer or both two layer. The only error in the assumed soil models is in the value of one of the soil parameters. For the single-layer models, we assign incorrect values to either $\sigma_1$ or $\epsilon_{r1}$. For the two-layer models, we assign incorrect values to $\sigma_1$, $\sigma_2$, $\epsilon_{r1}$, $\epsilon_{r2}$, or $L$, one at a time. The incorrect values correspond to values higher than those in Table~\ref{table_soil_parameters} used to compute $D$. We explore three levels of errors: $10\%$, $5\%$, and $1\%$. The same general approach was used by \citet{spinelli2022} to study the sensitivity of the LEDA experiment to soil parameter errors. 

Running a large number of FEKO simulations in which several parameters are varied simultaneously to robustly sample the soil parameter space is very computationally intensive. We leave this task to future work. We also leave to future work the exploration of errors in which the assumed number of soil layers is wrong.

\subsection{LST-averaging and Noise}
\label{section_lst_average_noise}
For each soil model, chromaticity correction, and error type tested, we study the Cosmic Dawn signal extraction after averaging the simulated spectra over $24$~hr of LST. Leveraging the time dependence of the beam-convolved foreground is expected to improve the constraints on the time-independent Cosmic Dawn signal \citep[e.g.,][]{liu2013,tauscher2020,anstey2023}. However, here we choose to work with the LST-averaged spectrum because (1) this choice simplifies our analysis, which represents an initial effort to quantify the effects from the soil and beam chromaticity; (2) averaging in LST reduces the measurement noise for a given observation time and is often a useful step when working with a limited amount of real data, motivating this approach in simulation; and (3) the variations with LST of the beam-convolved foreground observed from MARS are not as large as from lower latitudes (up to $\approx20\%$ in our simulations of the GSM), and, therefore, their leverage is expected to have a lower impact.

For each case being tested, the final simulated spectrum, $d(\nu)$, is produced by applying chromaticity correction to the data (with $C=1$ for the NC cases), averaging the data in LST, and adding random noise, that is,

\begin{align}
d(\nu) = \Biggl \langle \frac{T_S(\nu,t)}{C(\nu,t)} \Biggr \rangle + n(\nu),
\label{equation_average_spectrum}
\end{align}

\noindent where $\langle\ldots\rangle$ represents average over LST and $n(\nu)$ is the noise. For practicality, we add $n(\nu)$ in a single step after LST averaging the noiseless spectra. This is equivalent to LST averaging noise previously added to the $6$ minute cadence spectra. The noise has a Gaussian distribution, no correlation between frequency channels, and a standard deviation of $3$~mK at $75$~MHz, which evolves with frequency proportionally to the noiseless term. According to the radiometer equation, for a reference system temperature of $1500$~K, a channel width of $1$~MHz, and $100\%$ on-sky duty cycle, this noise level would be obtained with $2.9$ days of observations.

\subsection{Spectral Fitting}
\label{section_modeling}
We fit the following model to the simulated spectra:

\begin{equation}
m(\nu,\bm{\mu}) = \frac{T_{21}(\nu,\bm{\mu}_{21})}{\langle C(\nu,t)\rangle} + m_{fg}(\nu,\bm{\mu}_{fg}),
\label{equation_spectral_model}
\end{equation}

\noindent where $T_{21}$ is the model for the Cosmic Dawn signal. In this paper, we are interested in the recovery of the $21$~cm parameters assuming that the $21$~cm model itself is perfect. Therefore, $T_{21}$ in Equation~\ref{equation_spectral_model} is the Gaussian model of Equation~\ref{equation_global_model}, also used for the input $21$~cm signal. The $21$~cm fit parameters are $a_{21}$, $\nu_{21}$, and $w_{21}$, which are encapsulated by $\bm{\mu}_{21}$. Dividing $T_{21}$ by $\langle C(\nu,t) \rangle$ in Equation~\ref{equation_spectral_model} accounts for the effect of the beam chromaticity correction on the $21$~cm signal \citep{sims2023}. The variable $m_{fg}$ is a model for the beam-convolved, chromaticity-corrected, and LST-averaged foreground contribution, for which in this paper we use the ``LinLog'' expression \citep{hills2018,bradley2019,mahesh2021,sims2023}

\begin{equation}
m_{fg}(\nu,\bm{\mu}_{fg}) = \left(\frac{\nu}{\nu_{fg}}\right)^{-2.5}\sum_{i=0}^{N_{fg}-1} a_i \left[\log\left(\frac{\nu}{\nu_{fg}}\right)\right]^{i}.
\label{equation_foreground}
\end{equation}

\noindent Here, $\bm{\mu}_{fg}$ encapsulates the $a_i$ linear fit parameters, $N_{fg}$ is the number of terms in the expansion, and $\nu_{fg}=75$~MHz is a normalization frequency used to improve the numerical stability of the fit. We use the LinLog expression because it can efficiently model the MIST observations for different levels of chromaticity correction. We leave for future work the exploration of other analytical models, including different linear and nonlinear power law-based expansions \citep{voytek2014,bernardi2016,monsalve2017,bowman2018,hills2018,singh2022}.

In the simulated observations, the level of spectral structure from the beam-convolved, LST-averaged foreground is expected to vary across soil models and chromaticity correction cases. To optimize the fitting of this structure, we sweep $N_{fg}$ between $4$ and $8$, and keep for our analysis the results for the $N_{fg}$ that minimizes the Bayesian information criterion (BIC):

\begin{equation}
\mathrm{BIC}=\chi^2 + N_{\mu}\log N_{\nu},
\end{equation}

\noindent where $N_{\mu}=N_{fg}+3$ is the number of foreground plus $21$~cm fit parameters and $N_{\nu}=61$ is the number of frequency channels.

To fit the model parameters ---in particular $\bm{\mu}_{21}$--- in a computationally efficient way, we use the technique developed in \citet{monsalve2018}. In this technique, only the three nonlinear $21$~cm parameters are sampled, while the linear foreground parameters are fitted using the matrix operations of the linear least squares method. Despite only sampling the three-dimensional $\bm{\mu}_{21}$ space, the resulting probability density functions (PDFs) for $\bm{\mu}_{21}$ do account for covariances with the foreground parameters. We sample the $\bm{\mu}_{21}$ space using the \verb~pocoMC~ code, which implements the preconditioned Monte Carlo method for accelerated Bayesian inference \citep{karamanis2022, karamanis2023}. In the preconditioned Monte Carlo fit, we use the following uninformative uniform priors for the $21$~cm parameters: $[-1,1]$~K for $a_{21}$, $[45,105]$~MHz for $\nu_{21}$, and $[1,60]$~MHz for $w_{21}$. Our prior for $a_{21}$ is wide and would enable the detection of the input signal as well as of a large absorption feature produced by exotic physics \citep[e.g.,][]{bowman2018,feng2018,munoz2018}. More importantly for this paper, our prior for $a_{21}$ enables the detection of absorption and emission features corresponding to artifacts produced by soil effects combined with the limitations of our analysis approach. Although forcing the fitted $21$~cm model to be found in absorption could be warranted in some types of analyses, in this paper we choose to transparently expose the results affected by systematic effects. Our priors for $\nu_{21}$ and $w_{21}$ are chosen to match our frequency range.

\begin{figure*}
\centering
\includegraphics[width=\linewidth]{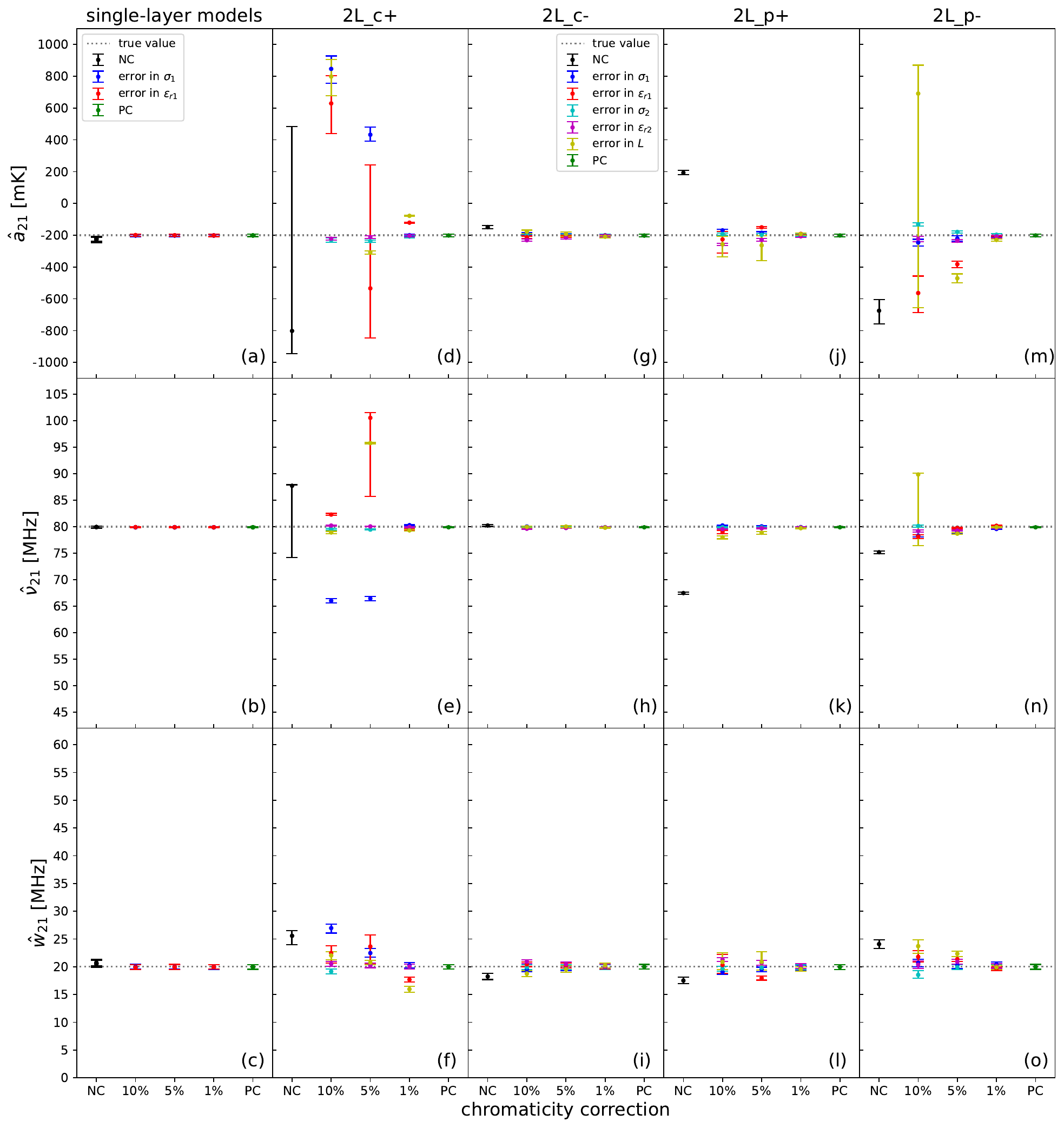}
\caption{Estimates for the parameters of our Gaussian $21$~cm absorption model, $a_{21}$, $\nu_{21}$, and $w_{21}$, obtained from fits to simulated observations with the MIST instrument over nine soil models. The first column shows the results for the five single-layer soil models. These results are similar and depicted with the same color for the same chromaticity correction case and error type. The second through fifth columns show the results for each of the two-layer models. On each panel, the $x$-axis indicates the different chromaticity correction cases. The results without chromaticity correction (NC) are depicted in black. The results with perfect chromaticity correction (PC) are shown in green. Other colors represent cases where the chromaticity correction has been applied but the value assumed for one of the soil parameters during the computation of the beam directivity has an error ($10\%$, $5\%$, or $1\%$). The legend for the two-layer soil  models is contained in panel~(g).}
\label{figure1}
\end{figure*}

\begin{figure*}
\centering
\includegraphics[width=\linewidth]{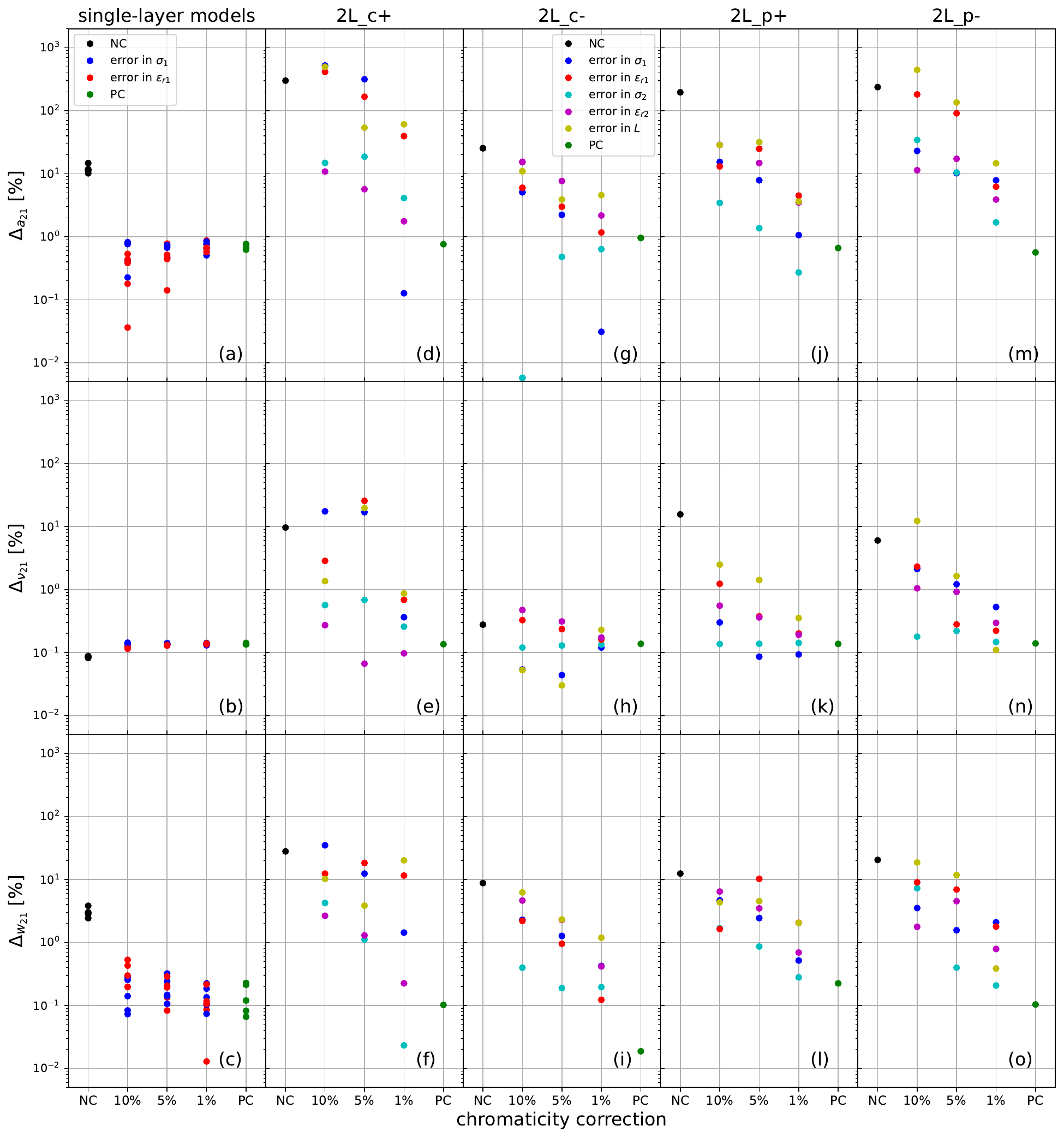}
\caption{Absolute percentage error of the estimates shown in Figure~\ref{figure1} computed using Equation~\ref{equation_error}.}
\label{figure2}
\end{figure*}

\begin{figure*}
\centering
\includegraphics[width=\linewidth]{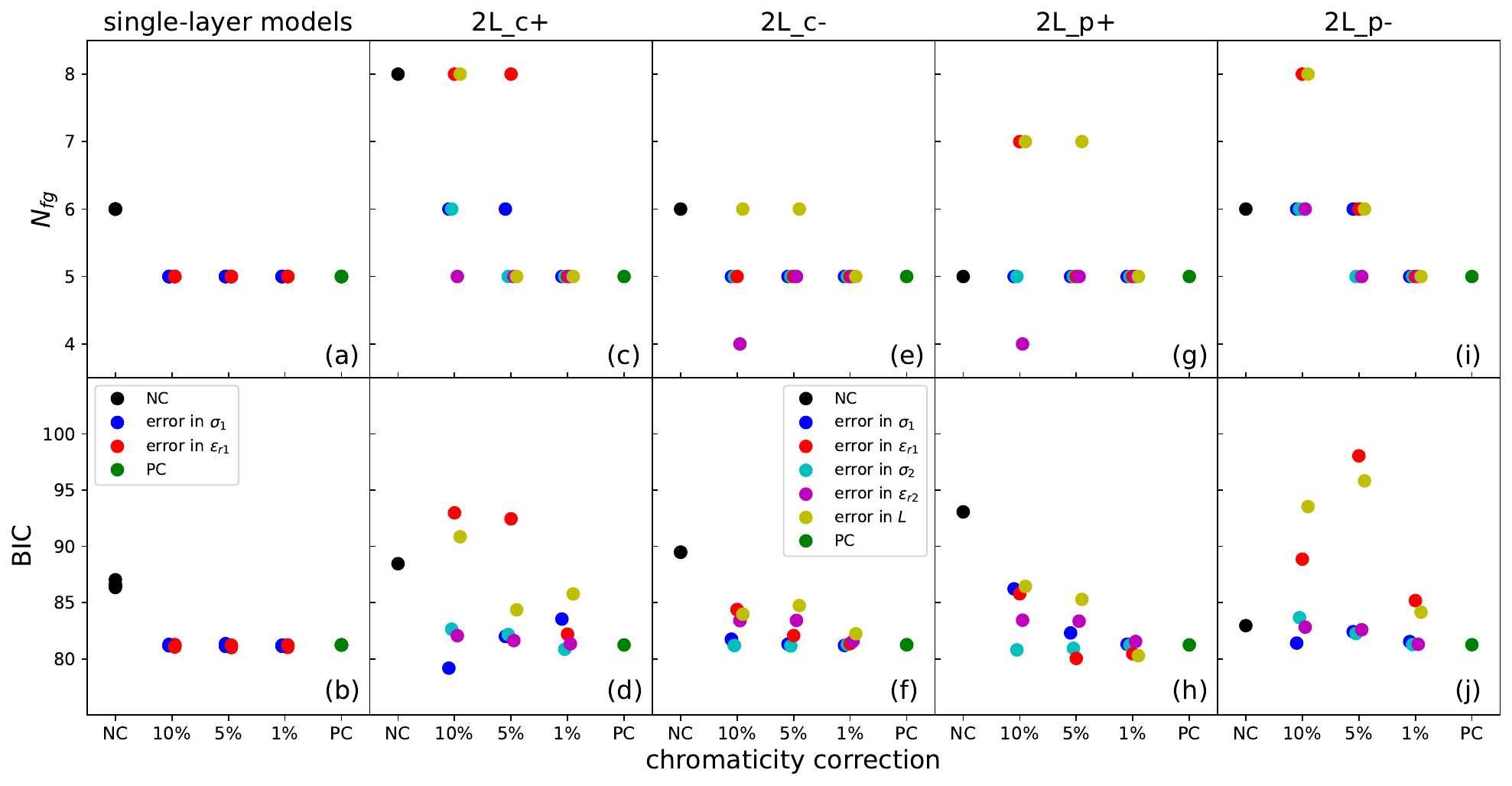}
\caption{(\textit{Top}) Number of terms in the LinLog foreground model used to produce the results of Figure~\ref{figure1}. These numbers were determined by minimizing the BIC. (\textit{Bottom}) BICs corresponding to the $N_{fg}$ in the top row. The reduced $\chi^2$ of the fits across all cases is in the range $0.81$--$1.17$, indicating that the LinLog model is reasonable and that there is no significant over- or underfitting.}
\label{figure3}
\end{figure*}

\section{Results and discussion}

The results for all the soil models, chromaticity corrections, and error types are presented in Figures~\ref{figure1}--\ref{figure3}.

Figure~\ref{figure1} shows the $21$~cm parameter estimates in terms of best-fit values and $1$-sigma\footnote{In this paper, the word ``sigma'' is used for the statistical uncertainty of the parameter estimates as the symbol $\sigma$ is used for the soil electrical conductivity.} error bars. The best-fit values correspond to the median or 50th percentile of the posterior PDFs. The ``$+$'' and ``$-$'' sigmas correspond to the 84th-50th and 16th-50th percentiles of the PDFs, respectively. 

Because of the low noise in our simulated observations, in most cases the $21$~cm estimates are limited by the accuracy of the best-fit value rather than by statistical precision. Therefore, as another performance metric, in Figure~\ref{figure2} we report the results in terms of the error of the best-fit value. Specifically, we report the absolute percentage error, $\Delta$, of each estimate. As an example, for $a_{21}$ this quantity is computed as

\begin{equation}
\Delta_{a_{21}}=100\% \times \left|\frac{\hat{a}_{21}-a_{21,\mathrm{input}}}
{a_{21,\mathrm{input}}}\right|,
\label{equation_error}
\end{equation}

\noindent where $\hat{a}_{21}$ is the best-fit value and $a_{21,\mathrm{input}}=-200$~mK is the input value.

Figure~\ref{figure3} shows the $N_{fg}$ used to fit the foreground contribution to the spectrum, as well as the BICs of the fits. The reduced $\chi^2$ corresponding to the BICs of the fits, computed as $\left(N_{\nu}-N_{\mu}\right)^{-1}\chi^2$, are in the range $0.81$--$1.17$. These values indicate that the LinLog model is reasonable across all cases and that there is no significant over- or underfitting.

The results for the five single-layer soil models are similar and shown in the first column of the figures. The second through fifth columns of the figures show the results for each of the four two-layer soil models.

\subsection{Perfect Chromaticity Correction}
We start the discussion of our results with the PC cases. In the PC cases, the impact from beam chromaticity has been perfectly removed and the corrected spectra represent observations conducted with beams that are effectively achromatic. Therefore, the PC results serve as the reference.

The $21$~cm estimates in the PC cases are obtained with errors below $1\%$ for $\hat{a}_{21}$, $0.2\%$ for $\hat{\nu}_{21}$, and $0.3\%$ for $\hat{w}_{21}$. These estimates are less than $2$~sigma away from the input values. For the nine soil models, the achromatic effective beams are different. As Figure~$7$ of \citet{monsalve2023} shows, the FWHM of the original beams at $\nu_r=75$~MHz (the frequency from which the achromatic effective beams are produced) spans the ranges [$57^{\circ}$, $63^{\circ}$] and [$103^{\circ}$, $121^{\circ}$] in the E and H~plane, respectively. The differences in the beams produce differences in the beam-convolved foreground contribution to the spectra. However, these differences do not lead to significant differences in the $21$~cm estimates, which remain consistent across soil models. Further, to model the beam-convolved foreground contribution, all the PC spectra need $N_{fg}=5$. In the absence of beam chromaticity, $N_{fg}=5$ quantifies the complexity of the spectra due to the foreground alone, in our case from the GSM. $N_{fg}=5$ is consistent with results from EDGES, in which five terms are needed to describe the measured spectra in the ranges $50$--$100$~MHz and $90$--$190$~MHz after applying chromaticity correction  \citep{monsalve2017,bowman2018}. The BICs of all the PC fits are $\approx81.24$. For $N_{\mu}=N_{fg}+3=8$, this BIC corresponds to a $\chi^2$ of $48.35$ and a reduced $\chi^2$ of $0.91$.

\subsection{Single-layer Models}
The figures show that there is high consistency in the $21$~cm estimates across our single-layer models. This consistency reflects that the estimates are not very sensitive to the absolute value of the beam directivity, which for our single-layer models can vary by more than $100\%$ at some angles and frequencies \citep{monsalve2023}. Further, the estimates closely match the input values, even without chromaticity correction. In the NC cases, the errors are $<20\%$ for $\hat{a}_{21}$, $0.1\%$ for $\hat{\nu}_{21}$, and $4\%$ for $\hat{w}_{21}$, and the estimates are less than $2$~sigma away from the input values. The NC spectra require $N_{fg}=6$. Although this value represents a higher spectral complexity than from the foreground alone, it is low compared to the number of terms used by other experiments when modeling real data without chromaticity correction. For instance, LEDA, measuring above soil and a $3$~m~$\times$~$3$~m ground plane, needed eight terms to model their spectrum over $50$--$85$~MHz \citep{bernardi2016}. SARAS~3, observing on a lake, needed seven terms over $55$--$85$~MHz \citep{singh2022}. The need for only six terms without chromaticity correction over $45$--$105$~MHz reflects the low beam chromaticity of MIST when observing over uniform soils. 

Applying a chromaticity correction significantly improves the accuracy and precision of the $21$~cm estimates, even when the assumed soil parameters have errors. Specifically, when the chromaticity correction is affected by errors in the soil parameters of $10\%$ or less, the errors in the $21$~cm estimates are comparable to the PC cases: $<1\%$ for $\hat{a}_{21}$, $0.2\%$ for $\hat{\nu}_{21}$, and $0.6\%$ for $\hat{w}_{21}$. Further, for these erroneous chromaticity corrections, $N_{fg}$ is $5$ and the BIC is almost the same as for the PC cases. These values for $N_{fg}$ and the BIC indicate that the correction errors do not significantly increase the complexity of the spectra above the intrinsic complexity of the foreground.

\subsection{Two-layer Models}
\label{section_two_layer}

As expected from the more complex directivity they produce, two-layer soil models make the extraction of the $21$~cm signal more difficult and, in general, impose requirements on the accuracy of the soil parameter values.

\subsubsection{Changes in Bottom-layer Conductivity}

Model \verb~2L_c+~ is the most challenging among our two-layer soil models because it produces the largest ripples in the simulated directivity \citep{monsalve2023}. Without chromaticity correction, the complexity in the spectrum produced by this soil model requires increasing $N_{fg}$ to $8$, as shown in panel~(c) of Figure~\ref{figure3}. The errors in the estimates are $\approx300\%$ for $\hat{a}_{21}$, $10\%$ for $\hat{\nu}_{21}$, and $30\%$ for $\hat{w}_{21}$. The large and uneven error bars seen in Figure~\ref{figure1}, in particular for $\hat{a}_{21}$ and $\hat{\nu}_{21}$, occur because in the NC case the posterior PDFs are wide and multimodal. When applying chromaticity correction with $10\%$ and $5\%$ errors in the soil parameters, the accuracy of the $21$~cm estimates does not improve compared to the NC case. In some of these cases, fitting the spectrum also requires an $N_{fg}$ as high as $8$, indicating that the imperfection of the correction leaves complex structure behind. The error bars of the estimates that require $N_{fg}=8$ are also large due to wide and multimodal PDFs. When the soil parameter errors in the chromaticity correction are reduced to $1\%$, the errors in $\hat{\nu}_{21}$ are reduced to $1\%$. However, the errors in $\hat{a}_{21}$ and $\hat{w}_{21}$ remain large: $\approx60\%$ and $20\%$, respectively. Therefore, conducting observations with MIST over the type of soil represented by model \verb~2L_c+~ would require a soil parameter accuracy better than $1\%$. Reaching this accuracy is anticipated to be extremely challenging and, hence, this type of soil should be avoided.

Model \verb~2L_c-~ enables the recovery of the $21$~cm parameters with the highest accuracy among our two-layer models. This occurs because the simulated directivity for \verb~2L_c-~ has the smallest ripples compared with the other two-layer models \citep{monsalve2023}. For this model, when chromaticity correction is not applied, the errors in the estimates are $\approx24\%$ for $\hat{a}_{21}$, $0.3\%$ for $\hat{\nu}_{21}$, and $9\%$ for $\hat{w}_{21}$, and the fitting of the spectrum requires $N_{fg}=6$. Applying chromaticity correction with increasing accuracy consistently reduces the errors in the $21$~cm estimates. In particular, with $5\%$ ($1\%$) errors in the soil parameters, the errors in $\hat{a}_{21}$ and $\hat{w}_{21}$ are reduced to $\approx8\%$ ($5\%$) and $2.3\%$ ($1.2\%$), respectively. Fitting the corrected spectra in most cases requires an $N_{fg}$ of $5$ or $6$. In one case, however, a value of $4$ is preferred. This case occurs when the correction is computed with a $10\%$ error in $\epsilon_{r2}$. Requiring an $N_{fg}<5$, as in this case, indicates that the spectral structure introduced by the imperfect chromaticity correction cancels out some of the structure in the intrinsic foreground spectrum.

\subsubsection{Changes in Bottom-layer Permittivity}

When the simulated observations are conducted over the \verb~2L_p+~ two-layer soil model, applying chromaticity correction is necessary for extracting the $21$~cm feature. Fitting the spectrum in the NC case requires $N_{fg}=5$. This value would initially suggest that the beam chromaticity does not significantly increase the complexity of the spectrum beyond that due to the foreground. However, in this case the Gaussian is found in emission instead of absorption and at an incorrect center frequency. The best-fit amplitude is $193$~mK (error of $196.5\%$) and the best-fit center is $67.5$~MHz (error of $15.6\%$). When chromaticity correction is applied, the Gaussian is found in absorption and the estimates cluster about the input parameter values. In particular, with $10\%$ errors in the soil parameters, the errors in the $21$~cm estimates are $30\%$ for $\hat{a}_{21}$, $2.4\%$ for $\hat{\nu}_{21}$, and $7\%$ for $\hat{w}_{21}$. Among our two-layer soil models, \verb~2L_p+~ follows \verb~2L_c-~ in terms of fidelity in the extraction of the Cosmic Dawn feature as long as chromaticity correction is applied. This result is consistent with the fact that \verb~2L_p+~ produces the second smallest ripples in the simulated directivity after \verb~2L_c-~ \citep{monsalve2023}.

For model \verb~2L_p-~, the extraction of the Cosmic Dawn feature is again challenging and requires chromaticity correction with a soil parameter accuracy of $1\%$. Without correction, the Gaussian is found in absorption but with large errors: $230\%$ for $\hat{a}_{21}$, $6\%$ for $\hat{\nu}_{21}$, and $20\%$ for $\hat{w}_{21}$. When correction is applied, a noticeable decrease in the errors of $\hat{\nu}_{21}$ and $\hat{w}_{21}$, to $1.6\%$ and $11\%$ respectively, is achieved if the errors in the soil parameters are reduced to $5\%$. However, sufficiently decreasing the error in $\hat{a}_{21}$, specifically to $14\%$, requires reducing the soil parameter errors to $1\%$. This is the second-strictest accuracy requirement among our models, consistent with \verb~2L_p-~ producing the second largest ripples in the simulated directivity following \verb~2L_c+~ \citep{monsalve2023}. Because of this tight accuracy requirement, this type of soil should also be avoided.

\section{Summary and future work}

In this paper, we use simulated observations to gain intuition about the effects of the MIST beam chromaticity on the detection of the global $21$~cm signal from the Cosmic Dawn. We attempt the detection of this signal for the nine models of the MIST beam directivity introduced in \citet{monsalve2023}. These directivity models come from electromagnetic simulations that incorporate different models for the soil. Five of the soil models are single-layer, uniform models. The other four are two-layer models in which either the electrical conductivity or relative permittivity changes below $1$~m from the surface. We study one phenomenological model for the Cosmic Dawn absorption feature. This model corresponds to a Gaussian with an amplitude of $-200$~mK, a center at $80$~MHz, and a full width at half maximum of $20$~MHz, consistent with standard physical models. 

Our five single-layer soil models yield accurate and precise estimates for the $21$~cm parameters even without making a correction for beam chromaticity before fitting the spectrum. Nonetheless, if chromaticity correction is applied, the estimates improve noticeably, becoming highly consistent with the input values even if the correction is imperfect due to errors in the parameters assumed for the soil. These results are very encouraging and strongly motivate us to optimize for soil uniformity when choosing an observation site. 

Two-layer soil models produce ripples in the beam directivity as a function of frequency, which make the extraction of the $21$~cm signal more challenging. The best results among our soil models are obtained for model \verb~2L_c-~, when the bottom soil layer has a lower conductivity than the top layer, which is the case that produces the smallest ripples. In this case, the $21$~cm parameters can be determined (with an accuracy that for the absorption amplitude is $24\%$ or better) even without chromaticity correction. The second best results are obtained for model \verb~2L_p+~, in which the bottom layer has a higher permittivity than the top layer and which produces larger ripples. In this case, however, chromaticity correction is required for the $21$~cm estimates to approach the input values. In models \verb~2L_p-~ and \verb~2L_c+~ the bottom layer has a lower permittivity and higher conductivity than the top layer, respectively, and the ripples in the directivity are even larger. In these two cases it is possible to recover the $21$~cm parameters with acceptable accuracy (in particular, an amplitude error within a few tens of a percent) only if chromaticity correction is applied with a soil parameter accuracy equal to and better than $1\%$, respectively. Meeting these requirements is expected to be extremely challenging; therefore, these two types of soils must be avoided.

Natural extensions to this analysis include the exploration of a wide range of models for the $21$~cm signal as well as for the soil. In the future, we will incorporate soil models with more than two layers and different layer thicknesses. We will also explore errors in the chromaticity correction due to assuming an incorrect number of layers for the soil. Sampling the soil parameter space in a statistically robust way can be computationally intensive, especially for soil models with two or more layers (i.e., five or more parameters), when each sample requires a new electromagnetic simulation of the instrument. To overcome this computational bottleneck, we are developing an emulator that will quickly generate realizations of the beam directivity after being trained on a sufficiently large set of precomputed electromagnetic simulations. This emulator will enable the efficient sampling of the soil parameter space and help reveal the covariances between soil and $21$~cm parameters. 

For each soil model, chromaticity correction, and error type tested, we studied the extraction of the Cosmic Dawn signal using a single spectrum. This spectrum corresponds to a $24$~hr average of observations simulated with a cadence of $6$ minutes from the latitude of MARS ($79.38^{\circ}$~N). To fit the foreground contribution to this spectrum we use the LinLog analytical model. In the cases with perfect chromaticity correction, this model requires five expansion terms, consistent with theoretical predictions and experimental results. The reduced $\chi^2$ of the PC fits is $0.91$ and across all cases is in the range $0.81$--$1.17$, which indicates that the LinLog model is a reasonable choice for this analysis. In the future we will explore other analytical models for the foreground contribution. Leveraging the time dependence of the beam-convolved foreground, in addition to improving the constraints on the time-independent $21$~cm signal \citep[e.g.,][]{liu2013,tauscher2020,anstey2023}, could relax the accuracy requirements on the soil parameters. We also leave the exploration of this possibility for future work.

\begin{acknowledgements}
We acknowledge support from ANID Chile Fondo 2018 QUIMAL/180003, Fondo 2020 ALMA/ASTRO20-0075, Fondo 2021 QUIMAL/ASTRO21-0053. We acknowledge support from Universidad Cat\'olica de la Sant\'isima Concepci\'on Fondo UCSC BIP-106. We acknowledge the support of the Natural Sciences and Engineering Research Council of Canada (NSERC), RGPIN-2019-04506, RGPNS 534549-19. This research was undertaken, in part, thanks to funding from the Canada 150 Research Chairs Program. This research was enabled in part by support provided by SciNet and the Digital Research Alliance of Canada. We acknowledge the support of the Canadian Space Agency (CSA) [21FAMCGB15]. We also acknowledge the Polar Continental Shelf Program for providing funding and logistical support for our research program, and we extend our sincere gratitude to the Resolute staff for their generous assistance and bottomless cookie jars.
\end{acknowledgements}

\software{Ipython \citep[][\url{https://ipython.org/}]{perez2007}, Numpy \citep[][\url{https://numpy.org/}]{harris2020}, Scipy \citep[][\url{https://doi.org/10.5281/zenodo.7502588}]{virtanen2020}, Matplotlib \citep[][\url{https://doi.org/10.5281/zenodo.7637593}]{hunter2007}, PyEphem \citep[][\url{https://pypi.org/project/ephem/}]{rhodes2011}, Astropy \citep[][\url{https://doi.org/10.5281/zenodo.6579729}]{astropy2013}, Healpy \citep[][\url{https://doi.org/10.5281/zenodo.6887117}]{gorski2005,zonca2019}, h5py (\url{https://doi.org/10.5281/zenodo.7568214}), pyGDSM \citep[][\url{https://doi.org/10.5281/zenodo.3835582}]{price2016}, pocoMC \citep[][\url{https://doi.org/10.5281/zenodo.7308533}]{karamanis2022, karamanis2023}}

\end{document}